\documentclass[prd,showpacs,twocolumn]{revtex4}

\begin{document}

\title{How to adapt broad-band gravitational-wave searches for $r$-modes}
\author{Benjamin J. Owen}
\affiliation{
Institute for Gravitation and the Cosmos,
Center for Gravitational Wave Physics,
Department of Physics,
The Pennsylvania State University, University Park, Pennsylvania 16802, USA
}
\affiliation{
Max Planck Institut f\"ur Gravitationsphysik (Albert Einstein Institut),
Callinstr.\ 38, 30167 Hannover, Germany
}

\date{$$Id: rpol.tex,v 1.32 2010/06/10 10:11:17 owen Exp $$}

\begin{abstract}

Up to now there has been no search for gravitational waves from the $r$-modes
of neutron stars in spite of the theoretical interest in the subject.
Several oddities of $r$-modes must be addressed to obtain an observational
result:
The gravitational radiation field is dominated by the mass current
(gravitomagnetic) quadrupole rather than the usual mass quadrupole, and the
consequent difference in polarization affects detection statistics and
parameter estimation.
To astrophysically interpret a detection or upper limit it is necessary to
convert the wave amplitude to an $r$-mode amplitude.
Also, it is helpful to know indirect limits on gravitational-wave emission to
gauge the interest of various searches.
Here I address these issues, thereby providing the ingredients to adapt
broad-band searches for continuous gravitational waves to obtain $r$-mode
results.
I also show that searches of existing data can already have interesting
sensitivities to $r$-modes.

\end{abstract}


\maketitle

\section{Introduction}

In recent years the LIGO Scientific Collaboration and Virgo Collaboration have
performed numerous searches for continuous gravitational-wave (GW) emission
from rapidly rotating neutron stars.
These include narrow-band searches for known pulsars~\cite{Abbott:2003yq,
Abbott:2004ig, Abbott:2007ce, Abbott:2008fx, Abbott:2009rf} (i.e.\ on or near a
timing solution), broad-band searches for the (non-pulsing) accreting neutron
star in the low-mass x-ray binary Sco~X-1~\cite{Abbott:2006vg, Abbott:2007tw},
and all-sky broad-band surveys for previously unknown neutron
stars~\cite{Abbott:2005pu, Abbott:2006vg, Abbott:2007tda, Abbott:2008uq,
Abbott:2008rg, Abbott:2009nc}.
Also, a broad-band search for the non-pulsing central compact object in
supernova remnant Cas~A is underway~\cite{Wette:2008hg}.

Cas~A is the youngest object targeted by a continuous GW search so far, and at
300 years old it may still be subject to the $r$-mode GW emission mechanism.
The $r$-modes~\cite{Papaloizou:1978} are rotation-dominated oscillations which
are driven unstable by GW emission~\cite{Andersson:1997xt, Friedman:1997uh}
(the ``Chandrasekhar-Friedman-Schutz'' version of the two-stream
instability~\cite{Chandrasekhar:1970, Friedman:1978}) even in the presence of
nuclear-matter viscosity~\cite{Lindblom:1998wf, Andersson:1998ze,
Lindblom:1999}.
This makes $r$-modes observationally a very interesting possibility for GW
emission in newborn neutron stars~\cite{Owen:1998xg} and accreting neutron
stars in low-mass x-ray binaries~\cite{Bildsten:1998ey, Andersson:1998qs}.
Nonlinear hydrodynamic saturation of the $r$-mode instability is now thought
(under most conditions) to occur at amplitudes several orders of magnitude
below those originally guessed~\cite{Arras:2002dw}, making unlikely the most
optimistic scenario of extragalactically detectable highly chirping signals
from young neutron stars~\cite{Owen:1998xg}.
However a low saturation amplitude would also keep $r$-modes active in neutron
stars up to thousands of years old under the right
conditions~\cite{Arras:2002dw}.
Therefore it is interesting to search for continuous nearly-periodic $r$-mode
GW emission from Cas~A and other very young pulsing or non-pulsing neutron
stars.

However up to now the continuous GW observational literature has assumed that
GW emission is from ``ellipticity,'' \textit{i.e.}\ an $\ell=m=2$ mass
quadrupole which rotates with the star.
By contrast $r$-modes are dominated by $\ell=m=2$ current quadrupole
emission~\cite{Lindblom:1998wf}.
The multipole structure of the radiation field affects the wave polarization
and detector response to a signal.
These in turn affect detection statistics and parameter estimation, most
notably the procedure used to convert the detector response $h(t)$ into an
intrinsic GW strain amplitude $h_0$ (or an observational upper limit on it).
Also for an $r$-mode the astrophysical interpretation of a measured $h_0$ or
upper limit must lead to a mode amplitude rather than an ellipticity.
(If the mode amplitude evolves in a complicated way, as predicted under some
conditions~\cite{Bondarescu:2008qx}, an observational upper limit would apply
to a time-rms amplitude.)
Finally, indirect limits on GW emission used as milestones for search
sensitivities and to evaluate potential searches are different for $r$-modes
than for ellipticity.

In this paper I address these amplitude- and polarization-related issues for
$r$-mode GW emission.
This is sufficient for broad-band searches, i.e.\ those not targeting a known
pulsar.
I do not address two important issues related to frequency and phase:
The $r$-mode frequency as a function of neutron star spin frequency depends on
the equation of state and relativistic effects (e.g.~\cite{Yoshida:2004gk}).
Frequency estimates are needed for GW searches targeted at known pulsars and
will be addressed elsewhere~\cite{JonesOwen}.
Also, in some scenarios~\cite{Bondarescu:2008qx} the $r$-modes may not maintain
phase coherence over likely observing times.
This also will be addressed elsewhere~\cite{Wasserman}.

The rest of this paper goes as follows:
In Sec.~\ref{multipoles}, I briefly review the basics of $r$-modes and the
multipole formalism of GW generation.
In Sec.~\ref{direct}, I show how these issues relate to GW observations,
translating the theory-oriented quantities in Sec.~\ref{multipoles} to those
used in the observational literature.
I present indirect limits on $r$-mode GW emission in Sec.~\ref{indirect}.
I summarize in Sec.~\ref{end}.
Unless otherwise noted I use geometrized units, where Newton's gravitational
constant and the speed of light are unity, and $1~M_\odot =
4.92\times10^{-6}$~s.
I also use ``neutron star'' in a broad sense which includes the possibility of
quark matter, meson condensates, and other exotica.

\section{Multipole structure}
\label{multipoles}

Here I describe ellipticity and $r$-mode GW emission in the context of the
multipole formalism for GW generation, particularly the canonical review by
\citet{Thorne:1980ru}.

The LIGO and Virgo continuous-wave observational literature characterizes the
mass quadrupole moment of a neutron star in terms of an equatorial ellipticity
\begin{equation}
\label{eps}
\epsilon = (I_{xx} - I_{yy}) / I_{zz},
\end{equation}
where the $z$-axis is the rotation axis of the star and $I_{ab}$ is the moment
of inertia tensor.
The latter is identical to the mass quadrupole tensor up to a trace, which is
not important here.
The $x$ and $y$ axes are chosen to co-rotate with the star so that $\epsilon$
does not oscillate.
It is convenient to relate $\epsilon$ to the scalar mass multipoles $I^{\ell,
m}$, which are given in a non-rotating frame and the Newtonian limit
by~\cite{Thorne:1980ru}
\begin{equation}
I^{\ell,m} = \frac{16\pi} {(2\ell+1)!!} \sqrt{\frac{(\ell+1)(\ell+2)}
{2\ell(\ell-1)}} \int d^3r \rho r^\ell Y^{\ell,m*},
\end{equation}
with $\rho$ representing the mass density, $r$ the radial coordinate, and
$Y^{\ell,m}$ the standard spherical harmonic.
For a real-valued density perturbation fixed with respect to a star rotating
with angular velocity $\Omega$, the time dependence of these is $I^{\ell,m} =
I^{\ell,-m*} \propto e^{-im\Omega t}$.
From e.g.\ Eq.~(4.7a) of Ref.~\cite{Thorne:1980ru} we obtain the relation
\begin{equation}
\left| I^{2,2} \right| = \sqrt{\frac{8\pi}{5}} I_{zz} \epsilon.
\label{itrans}
\end{equation}
This can be checked by examining the GW luminosity:
Inserting Eq.~(\ref{itrans}) into Eq.~(4.16) of Ref.~\cite{Thorne:1980ru}
obtains, e.g., Eq.~(20) of Ref.~\cite{Abbott:2006vg}:
\begin{equation}
\frac{dE}{dt} = \frac{\omega^6}{16\pi} \left| I^{2,2} \right|^2 =
\frac{\omega^6}{10} I_{zz}^2 \epsilon^2,
\label{dEdtI}
\end{equation}
where $\omega$ is the angular frequency of the wave, equal to $m\Omega$ for a
perturbation rotating with the star.
(The above expressions assume a negligible contribution from all mass
multipoles except $I^{2,2} = I^{2,-2*}$, but there are arguments that $I^{2,1}$
also can be significant in some stars---see \citet{Jones2010} for a new one and
a summary of older ones.)

In Newtonian gravity and the slow-rotation approximation, an $r$-mode is an
Eulerian velocity perturbation~\cite{Provost:1981}
\begin{equation}
\delta v_j = \alpha \Omega R (r/R)^\ell Y_j^{B,\ell,\ell} e^{i\omega t},
\label{deltav}
\end{equation}
where $\alpha$ is a dimensionless amplitude, $R$ is the stellar radius, and the
magnetic-parity vector spherical harmonic is defined in terms of scalar
spherical harmonics $Y^{\ell,m}$ as~\cite{Thorne:1980ru}
\begin{equation}
Y_j^{B,\ell,m} = [\ell(\ell+1)]^{-1/2} \epsilon_{jkp} rN^k \nabla^p Y^{\ell,m},
\end{equation}
where $N^k$ is the unit vector pointing from the center of the star.
The GW frequency $f=2\pi/\omega$ is identical to the mode frequency, which is
related to the spin frequency by~\cite{Papaloizou:1978}
\begin{equation}
\label{f}
\omega = -\frac{(\ell+2)(\ell-1)}{\ell+1} \Omega
\end{equation}
(using $\ell=m$ for the proper $r$-modes).

The GW strain tensor can be expanded as~\cite{Thorne:1980ru}
\begin{equation}
\label{hjk}
h_{jk} = \frac{1}{r} \sum_{\ell=2}^\infty \sum_{m=-\ell}^{+\ell} \left(
\frac{d}{dt} \right)^\ell \left( I^{\ell,m} T_{jk}^{E2,\ell,m} + S^{\ell,m}
T_{jk}^{B2,\ell, m} \right),
\end{equation}
where the spin-2 tensor spherical harmonics of electric (E) and magnetic (B)
parity are defined~\cite{Thorne:1980ru}
\begin{eqnarray}
T_{jk}^{E2,\ell,m} &=& \sqrt{\frac{2(\ell-2)!} {(\ell+2)!}} r^2 \nabla_j
\nabla_k Y^{\ell,m} - \mathrm{trace},
\\
T_{jk}^{B2,\ell,m} &=& N_p T_{q(j}^{E2,\ell,m} \epsilon_{k)pq}.
\label{BEidentity}
\end{eqnarray}
As well as the mass multipoles $I^{\ell,m}$ we encounter the gravitomagnetic or
current multipoles
\begin{equation}
\label{Slm}
S^{\ell,m} = \frac{-32\pi} {(2\ell+1)!!} \sqrt{\frac{\ell+2} {2(\ell-1)}} \int
d^3r \rho r^\ell v_j Y_j^{B,\ell,m*}.
\end{equation}
The time derivatives in Eq.~(\ref{hjk}) suppress GW emission from higher-$\ell$
multipoles by powers of a characteristic velocity, and thus the $\ell=2$
multipoles tend to be dominant.
For $r$-modes, the $\ell=2$ mode is also the most unstable and the least damped
by viscosity~\cite{Lindblom:1998wf}.

It is also useful to write the full GW luminosity including all
multipoles~\cite{Thorne:1980ru}
\begin{equation}
\frac{dE}{dt} = \frac{1}{32\pi} \sum_{\ell=2}^\infty \sum_{m=-\ell}^{+\ell}
\langle \left| \left( \frac{d}{dt} \right)^{\ell+1} I^{\ell,m} \right|^2 +
\left| \left( \frac{d}{dt} \right)^{\ell+1} S^{\ell,m} \right|^2 \rangle,
\label{dEdt}
\end{equation}
where the angle brackets denote a time average.

By using Eq.~(\ref{Slm}) and orthonormality relations from
\citet{Thorne:1980ru}, we see that an $\ell=2$ $r$-mode produces a current
quadrupole~\cite{Owen:1998xg}
\begin{equation}
S^{2,2} = -\frac{32\sqrt{2}\pi} {15} \alpha M \Omega R^3 \tilde{J} e^{i\omega
t},
\label{S22}
\end{equation}
where $\tilde{J}$ is a dimensionless functional of the neutron-star equation of
state and $M$ and $R$ are the mass and radius of the (unperturbed) star.
Neglecting the $r$-mode density perturbation, all multipoles other than
$S^{2,2}$ vanish---including $I^{2,2}$ and $I^{2,-2}$.

The $r$-mode amplitude $\alpha$ which I use here is related to others as
follows:
The results of \citet{Arras:2002dw} are expressed in terms of $A_1 = \alpha
\sqrt{\tilde{J}/2}$, sometimes called $c_\alpha$ in other papers, which is
about $0.1\alpha$ for fiducial neutron star parameters (see below).
\citet{Bondarescu:2008qx} express most of their results in terms of
$\bar{C}_\alpha$, which removes the adiabatic change of amplitude as the star
spins down.
For the example mode triplet they use, the zero-viscosity parametric
instability threshold is normalized to $|C_\alpha|_0 = 0.011$~s$^{-1/2}$.
This makes their $\bar{C}_\alpha$ amplitudes about $1.1\times10^3 \alpha
\sqrt{\Omega/8.4\times10^3\mbox{ s}^{-1}}$ or about 200$\alpha$ at a spin
frequency of 100~Hz.

In Eqs.~(\ref{deltav}) and (\ref{f}) and the others derived from them, I have
neglected corrections of relative order $\Omega^2/(\pi \bar\rho)$ where
$\bar\rho$ is the mean density of the non-rotating star.
These include the density perturbation associated with an $r$-mode, which
contributes to the GW tensor and mass quadrupole also at relative order
$\Omega^2/(\pi\bar\rho)$.
These corrections rise to the 10\% level only for stars rotating faster than
the fastest known pulsars~\cite{Lindblom:1999}.
In view of the uncertain factors of 2 or more in quantities such as the moment
of inertia (see below), these and the somewhat larger corrections due to
relativistic gravity~\cite{Lockitch:1998nq} are not important.
The exception is the correction to Eq.~(\ref{f}), which affects narrow-band
searches for known pulsars and will be addressed elsewhere~\cite{JonesOwen}.

\section{Direct observations}
\label{direct}

Here I discuss the effects of $r$-mode multipole structure on aspects of direct
GW observations, including polarizations, detection statistics, waveform
parameter estimation (especially amplitude), and upper limit procedures.

Equation~(\ref{hjk}) implies that GWs from $r$-modes will have different
polarizations than GWs from ellipticity.
To see this compare two GWs with equal luminosity~(\ref{dEdt}), one dominated
by a mass multipole $I^{\ell,m}$ and one dominated by the corresponding current
multipole $S^{\ell,m}$.
The linear $+$ and $\times$ polarizations are projected out of the GW strain
tensor as
\begin{eqnarray}
h_+ &=& h_{jk} \left( p^jp^k - q^jq^k \right)/2 = h_{jk} e_+^{jk},
\nonumber
\\
h_\times &=& h_{jk} \left( p^jq^k + q^jp^k \right)/2 = h_{jk} e_\times^{jk},
\label{hpc}
\end{eqnarray}
where the unit vectors $p^j$ and $q^j$ are orthogonal to each other and to
$N^j$.
Using the identity $q_j = \epsilon_{jab} N_a p_b$ with Eq.~(\ref{BEidentity}),
we find that
\begin{eqnarray}
T_{jk}^{B2,\ell,m} e_+^{jk} &=& - T_{jk}^{E2,\ell,m} e_\times^{jk},
\nonumber
\\
T_{jk}^{B2,\ell,m} e_\times^{jk} &=& T_{jk}^{E2,\ell,m} e_+^{jk}.
\label{Ttrans}
\end{eqnarray}
For a fixed luminosity, the GW tensor $h_{jk}$ of the current-dominated GW is
obtained from that of the mass-dominated signal by taking $T_{jk}^{E2,\ell,m}
\to T_{jk}^{B2,\ell,m}$.
Substituting Eq.~(\ref{hjk}) into Eq.~(\ref{hpc}) and combining with
Eq.~(\ref{Ttrans}), we see that this takes
\begin{equation}
(h_+,h_\times) \to (-h_\times,h_+).
\label{trans}
\end{equation}
Since the $+$ and $\times$ polarizations are positive and negative parity
eigenstates, and the difference between mass and current multipoles is parity,
this was to be expected.
Note that the GW power radiated at a given angle from the star's rotation axis
is proportional to $h_+^2 + h_\times^2$---e.g.\ Eqs.~(4.12) and~(4.13) of
\citet{Thorne:1980ru}---and thus is not affected.

The strain response of a detector is the contraction of the GW tensor $h_{jk}$
with a tensor describing the detector's beam pattern.
It can be written by combining Eqs.~(10)--(11) and (20)--(22) of
Ref.~\cite{Jaranowski:1998qm} as (making the time dependence explicit)
\begin{eqnarray}
h(t) &=& [a(t) \cos2\psi + b(t) \sin2\psi] h_+(t)
\nonumber
\\
&& + [b(t) \cos2\psi - a(t) \sin2\psi] h_\times(t),
\label{hoft}
\end{eqnarray}
where $\psi$ is the polarization angle, explained simply in footnote 4 of
Ref.~\cite{Abbott:2003yq}, whose definition is equivalent to fixing $p^j$ and
$q^j$ in Eq.~(\ref{hpc}).
(The precise forms of the modulation functions $a$ and $b$ arising from the
rotation of the Earth-based detector are not needed here.)
From this we see that Eq.~(\ref{trans}) is equivalent to taking
\begin{equation}
\psi \to \psi + \pi/4.
\label{trans2}
\end{equation}
The transformation~(\ref{trans2}) lets us quickly examine the suitability for
$r$-modes of data analysis methods developed for ellipticity.

Many GW search methods assume a uniform Bayesian prior on $\psi$ and thus are
not affected by the transformation~(\ref{trans2}).
The $\mathcal{F}$-statistic of \citet{Jaranowski:1998qm} is one, as can be seen
by the lack of $\psi$ in their Eq.~(55) and its derivation.
The $\mathcal{F}$-statistic was derived as a frequentist maximization over
$\psi$ (and other angles); but when deriving a Bayesian alternative
$\mathcal{B}$-statistic, \citet{Prix:2009tq} explicitly showed that
$\mathcal{F}$ has an implicit uniform prior on $\psi$.
The $\mathcal{B}$-statistic itself has an explicit uniform prior on $\psi$.
The multi-interferometer $\mathcal{F}$-statistic~\cite{Cutler:2005hc} performs
a weighted sum of $\mathcal{F}$ over detectors and thus also has an implicit
uniform prior in $\psi$.
The heterodyning method used in most known-pulsar GW searches is Bayesian and
usually uses an explicit uniform prior in $\psi$, as in Eq.~(15) of
\citet{Dupuis:2005xv}.
Semi-coherent searches for unknown pulsars based on combining the raw
power~\cite{Abbott:2005pu, Abbott:2007tda, Abbott:2008rg} or
$\mathcal{F}$-statistics~\cite{Abbott:2008uq, Abbott:2009nc} of short stretches
of data also effectively use implicit uniform priors on $\psi$ (see
\citet{Abbott:2007tda}, especially the appendices, for details of the former).
The radiometer search for Sco~X-1~\cite{Abbott:2007tw} was adapted from a
stochastic background search and explicitly assumes no preferred polarization.

Some recent known-pulsar searches~\cite{Abbott:2008fx, Abbott:2009rf} use
values of $\psi$ from measurements of pulsar wind nebulae.
The measurement and error estimates can be folded into a Bayesian method
\cite{Dupuis:2005xv} as non-uniform priors on $\psi$, or the best $\psi$ value
can be inserted into the $\mathcal{F}$-statistic to obtain the
$\mathcal{G}$-statistic \cite{Jaranowski:2010rn}.
In order to cover the possibility of $r$-modes, such known-pulsar searches will
need to target not only different GW frequencies from the ellipticity case but
also $\psi \to \psi + \pi/4$.

Parameter estimation of candidate signals is also affected by
Eq.~(\ref{trans2}).
The observational literature characterizes the continuous GW amplitude with an
intrinsic strain amplitude defined in terms of ellipticity as
\begin{equation}
h_0 = r^{-1} \omega^2 I_{zz} \epsilon = r^{-1} \omega^2 \sqrt{\frac{5}{8\pi}}
\left| I^{2,2} \right|,
\end{equation}
where the second equality uses Eq.~(\ref{itrans}).
This $h_0$ is the amplitude of the response of a hypothetical detector at
either of the Earth's poles to a signal originating from a star over either
pole whose rotation axis is parallel to that of the Earth.
It is also simply related to the GW luminosity.
The full parameter estimation problem is lengthy and I do not address it here,
but I note the following simple and useful approximation.
A detected signal will be integrated for much more than one day, since even
semi-coherent search candidates will be followed up coherently.
In this limit \citet{Jaranowski:1998ge} performed detailed simulations
confirming that the beam-pattern modulation averages out and the signal is
effectively replaced by an unmodulated sine wave with amplitude
\begin{equation}
h_\mathrm{eff} = h_0 \sqrt{A + B\cos(4\psi)},
\end{equation}
where the lengthy expressions for the functions $A$ and $B$ are given in their
Eqs.~(36) \textit{et seq}.
Thus $\psi \to \psi + \pi/4$ simply flips the sign of $B$, which for most most
source parameters is a few percent correction to $h_\mathrm{eff}$.
(It may seem strange that $h_0$ is affected; this is because we are no longer
transforming $\psi$ at fixed luminosity but rather for fixed detector
response.)

An accurate measurement of $\psi$ could yield information on whether a signal
is from ellipticity or an $r$-mode:
For a known pulsar the ratio of GW frequency to spin frequency already
distinguishes between mechanisms and thus $\psi$ is redundant.
But if the GW signal comes from a pulsar wind nebula without a timed pulsar,
and the GW-measured $\psi$ (assuming ellipticity) is inconsistent with the
orientation of a jet or torus, this would indicate $r$-mode emission.
Also, a GW signal coming from Sco~X-1 or another non-pulsing accreting neutron
star could be compared with possible orientations from radio jets
\citep{Fomalont2001} or x-ray reprocessing \citep{Munoz-Darias2007} (since the
difference between these estimates is less than $\pi/4$).

Upper limits on $h_0$ from the uniform-$\psi$ searches are obtained in terms of
populations of software-injected signals uniformly distributed in $\psi$, and
hence the upper limit procedures for ellipticity remain valid for $r$-modes.
Upper limits on $h_0$ from searches with a given $\psi$ determined e.g.\ by a
pulsar wind nebula will need to use populations of injections taking
Eq.~(\ref{trans2}) into account to produce separate limits for ellipticity and
$r$-modes.

To convert an estimated or upper-limit $h_0$ to an $r$-mode amplitude, it is
convenient to write the GW luminosity as
\begin{equation}
\frac{dE}{dt} = \frac{1}{10} \omega^2 r^2 h_0^2
\label{dEdth0}
\end{equation}
and compare it to Eq.~(\ref{dEdt}),
\begin{equation}
\frac{dE}{dt} = \frac{\omega^6}{32\pi} \left| S^{2,2} \right|^2
\label{dEdtS}
\end{equation}
if $S^{2,2}$ is the only non-vanishing multipole.
[Note that the numerical coefficient is 1/2 that of Eq.~(\ref{dEdtI}) because
the $r$-mode is traditionally defined as a complex perturbation while the
ellipticity is defined as real, and $e^{i\omega t}$ contains twice as much
power as $\cos(\omega t)$ which lacks the imaginary part.]
Equating~(\ref{dEdth0}) and~(\ref{dEdtS}) obtains
\begin{equation}
h_0 = \sqrt{\frac{8\pi}{5}} r^{-1} \omega^3 \alpha M R^3 \tilde{J}.
\label{h0alpha}
\end{equation}
This equation allows us to convert $h_0$ to $\alpha$ for a fiducial value of
$MR^3\tilde{J}$, in the same way that $h_0$ is converted to $\epsilon$ for a
fiducial value of $I_{zz} = 10^{45}$~g\,cm$^2$.
(In both cases what is really measured is a quadrupole---and, if the star's
rotation axis is known, the parity which determines if it is a mass or current
quadrupole.)
Much of the theoretical neutron star literature uses a polytropic equation of
state with polytropic index $n=1$, which (for Newtonian gravity) fixes
$\tilde{J} \approx 0.0164$~\cite{Owen:1998xg}.
Combined with the above choice of $I_{zz}$ and the usual choice of
$M=1.4~M_\odot$, this fixes $R\approx11.7$~km rather than the common choices of
10~km and 12.53~km.
Inverting Eq.~(\ref{h0alpha}) and substituting the fiducial neutron star
values above, we obtain
\begin{equation}
\alpha = 0.028 \left( \frac{h_0}{10^{-24}} \right) \left( \frac{r}{\mbox{1
kpc}} \right) \left( \frac{\mbox{100 Hz}}{f} \right)^3.
\end{equation}
Note that modern investigations of mode saturation~\cite{Bondarescu:2008qx}
indicate that $\alpha$ may vary substantially over a typical observing time.
Since detection statistics respond to integrated signal power, the $\alpha$
inferred from them is really a time-rms average.

Here and below, when I give numerical values I assume the fiducial
neutron-star parameters above and I do not write scalings with $M$, $R$,
$I_{zz}$, and $\tilde{J}$.
That is because, even if the equation of state (EOS) is not specified, as long
as it is barotropic only two of the quantities are independent.
By noting that $I_{zz}$ can be written as $MR^2$ times a relatively
EOS-independent function (e.g.\ \citet{Lattimer:2000nx}) except for very
low-mass stars, and that $\tilde{J}$ is less EOS-dependent than $MR^3$, I
estimate that the EOS-related uncertainty in these quantities is dominated by
uncertainties in $M$ and $R$ and is therefore usually a factor 2--3.

\section{Indirect limits}
\label{indirect}

Indirect limits on GW emission inferred from electromagnetic astronomical
observables are useful to gauge the astrophysical interest of existing and
planned GW searches.
As in Ref.~\cite{Owen:2009tj}, where I summarize this and related issues in
more detail, I divide the sources into the four categories of accreting neutron
stars, known pulsars and non-pulsing stars without accretion, and the large
population of unseen neutron stars sought by all-sky surveys.

For known pulsars the primary indirect limit is the ``spin-down limit,''
obtained by assuming that all of the observed change in spin frequency is due
to GW emission.
Thus Eq.~(\ref{dEdth0}) is equated to the kinetic energy loss
\begin{equation}
\frac{dE}{dt} = I_{zz} \Omega \dot\Omega \approx (9/16) I_{zz} \omega
\dot\omega
\end{equation}
to obtain
\begin{equation}
h_0^\mathrm{sd} = \frac{1}{r} \sqrt{ \frac{45I_{zz}\dot{P}} {8P} }
\label{h0sd}
\end{equation}
where $P=2\pi/\Omega$ is the observed pulse period.
For $r$-modes this is 3/2 times the value for ellipticity-dominated emission,
given for instance by Eq.~(2) of \citet{Wette:2008hg}.
Substituting Eq.~(\ref{h0alpha}) obtains
\begin{eqnarray}
\alpha_\mathrm{sd} &=& \frac{405}{4096\pi^{7/2}} \sqrt{I_{zz} \dot{P}P^5}
M^{-1} R^{-3} \tilde{J}^{-1}
\nonumber
\\
&=& 1.4 \left( \frac{P}{\mbox{10 ms}} \right)^{5/2} \left(
\frac{\dot{P}}{10^{-10}} \right)^{1/2}
\nonumber
\\
&=& 0.033 \left( \frac{\mbox{100 Hz}} {f} \right)^{7/2} \left( \frac{|\dot{f}|}
{10^{-10}\mbox{ Hz\,s}^{-1}} \right)^{1/2},
\label{asd}
\end{eqnarray}
where the last equality uses $f=4/3/P$.
Again these are time-rms $\alpha$ limits if $\alpha$ fluctuates over the
observation time as possible in many young neutron star scenarios
\citep{Bondarescu:2008qx}.
Based on numbers from the on-line version 1.40 of the ATNF
catalog~\cite{Manchester:2004bp}, the pulsars with dipole spin-down ages
$P/(2\dot{P})$ less than $10^4$~yr generally have $\alpha_\mathrm{sd}$ of
order unity, except for J0537$-$6910 which has about 0.1.
\citet{Reisenegger:2003cq} also pointed out that, under the right conditions of
cooling and viscosity, $r$-modes could remain active in millisecond pulsars for
a long time (comparable to the spin-down age) after the star stops accreting as
a low-mass x-ray binary.
The millisecond pulsars typically have $\alpha_\mathrm{sd}$ of order
$10^{-6}$--$10^{-4}$, with a few as high as order $10^{-2}$.

A numerically stricter but more model-dependent indirect limit was obtained by
\citet{Palomba:1999su} for a few known pulsars by incorporating additional
information on the age and braking index
\begin{equation}
n = \Omega\ddot{\Omega} / \dot{\Omega}^2.
\end{equation}
If $\dot{\Omega}$ is proportional to a power of $\Omega$, the braking index
picks out that power.
For radiative braking from a static (electromagnetic or GW) multipole $\ell$,
the power is $n=2\ell+1$, while for particle winds it can be lower.
The highest value usually considered is $n=7$, which corresponds to constant
$\alpha$.
As first argued from adiabatic invariance~\cite{Ho:1999fh} and later seen
explicitly in nonlinear hydrodynamic saturation
calculations~\cite{Bondarescu:2008qx}, the long-term average of $\alpha$ might
scale as $\Omega^{-1/2}$, resulting in $n=6$.
\citet{Palomba:1999su} took $\dot{\Omega}$ to be a sum of two powers of
$\Omega$, one power 5 (GWs from ellipticity) and one a free parameter; and
performed numerical spin-down evolutions of the pulsars over a wide parameter
space constrained to be consistent with their known ages and present values of
$\Omega$, $\dot{\Omega}$, and $n$.
Because all observed braking indices are less than 3, the $\Omega^5$ GW
component of $\dot{\Omega}$ is constrained to be less than the spin-down limit
by some factor---in the case of the Crab pulsar, this limit on $h_0$ is 2.5
times stricter than the spin-down limit.
Without performing such detailed simulations, it is clear that since $r$-modes
have $\dot\Omega$ proportional to higher powers of $\Omega$, this type of limit
on the Crab would be stricter than the spin-down limit by more than the factor
2.5 for ellipticity.

Compare $\alpha_\mathrm{sd}$ to the theory of $r$-mode non-linear
hydrodynamical saturation:
The lowest zero-viscosity parametric instability threshold ($\bar{C}_\alpha=1$
in the notation of \citet{Bondarescu:2008qx}) corresponds to $\alpha$ a few
times $10^{-3}$ for the frequencies of interest.
The real threshold, which tends to serve as an attractor for $r$-mode
evolutions, depends on temperature as well as frequency and can be an order of
magnitude higher (e.g.\ their Fig.~6).
But it is still below the values of $\alpha_\mathrm{sd}$ for young pulsars,
which are high enough to lie in the ``run-away'' regime of non-linear
hydrodynamics which requires further study~\cite{Bondarescu:2008qx}.
For millisecond pulsars, the appropriate comparison is to lower values (see the
discussion of accreting stars below).

For non-accreting neutron stars without observed spins, such as the central
compact object in Cas~A, age-based indirect limits are obtained by substituting
the age of the object into the spin-down limits under the assumption that the
object has spun down significantly and predominantly by GW
emission~\cite{Wette:2008hg}.
In this case we have the relation
\begin{equation}
P/\dot{P} = (n-1)a,
\label{n}
\end{equation}
where $a$ is the age of the neutron star and the braking index $n$ is assumed
to be constant or appropriately averaged.
In parallel to the derivation of the ellipticity version of this
limit~\cite{Wette:2008hg}, we substitute Eq.~(\ref{n}) into Eq.~(\ref{h0sd}) to
obtain $\sqrt{3/2}$ the value for ellipticity, or
\begin{equation}
h_0^\mathrm{age} = 1.5\times10^{-24} \left( \frac{\mbox{300 yr}} {a}
\right)^{1/2} \left( \frac{\mbox{3.4 kpc}} {r} \right) \left( \frac{6}{n-1}
\right)^{1/2}.
\end{equation}
(The fiducial values are for Cas~A and constant-$\alpha$ evolution.)
Similarly for the age-based indirect limit on $r$-mode amplitude we obtain
\begin{eqnarray}
\alpha_\mathrm{age} &=& \frac{15}{8\sqrt{\pi}} \tilde{J}^{-1} I^{1/2} M^{-1}
R^{-3} \omega^{-3} a^{-1/2} (n-1)^{-1/2}
\nonumber
\\
&=& 0.14 \left( \frac{\mbox{300 yr}}{a} \right)^{1/2} \left( \frac{\mbox{100
Hz}}{f} \right)^3 \left( \frac{6}{n-1} \right)^{1/2},
\end{eqnarray}
where again the limit is a time-rms value if $\alpha$ fluctuates quickly
compared to the spin-down timescale.
For Cas~A, the youngest known neutron star, the values of $\alpha_\mathrm{age}$
are 0.14--0.005 over the 100--300~Hz band considered by \citet{Wette:2008hg}.
Several similar objects, e.g.\ the non-pulsing ones listed by
\citet{Halpern:2009bj}, have indirect limits of the same order of magnitude.
At a few hundred~Hz these values are comparable to the parametric instability
thresholds in several scenarios analyzed by \citet{Bondarescu:2008qx}, with
zero-viscosity $\alpha$ thresholds a few times $10^{-3}$ and true thresholds
several times higher.

For accreting neutron stars the standard indirect limit is derived from the
argument that accretion torque and GW torque are in equilibrium, originally
made by \citet{Papaloizou1978b}, put on firm astrophysical footing by
\citet{Wagoner:1984pv}, and later tied to observations of spins of accreting
neutron stars~\cite{Bildsten:1998ey, Andersson:1998qs}.
From $dJ/dt = (2/\omega) dE/dt$---as in Eq.~(4.23) of
\citet{Thorne:1980ru}---we have
\begin{equation}
h_0^2 = 5 \omega^{-1} r^{-2} dJ/dt,
\end{equation}
equivalent to Eq.~(4) of \citet{Watts:2008qw}.
The indirect limit on $h_0$ is obtained by assuming accretion of Keplerian
angular momentum at the stellar surface and 100\% conversion of gravitational
potential energy to x-ray flux $F_x$ measured at Earth to obtain
\begin{eqnarray}
h_0^\mathrm{acc} &=& \sqrt{20\pi} \omega^{-1/2} F_x^{1/2} M^{-1/4} R^{3/4}
\nonumber
\\
&=& 3.3\times10^{-26} \left( \frac{F_x} {3.9\times10^{-7}\mbox{ erg\,cm}^{-2}}
\right)^{1/2}
\nonumber
\\
&& \times
\left( \frac{\mbox{800 Hz}}{f} \right)^{1/2}.
\end{eqnarray}
Here the numerical value is scaled to the average bolometric flux of Sco~X-1
(which does not pulse) and the highest spin rate observed in an accreting
millisecond pulsar, both from~\citet{Watts:2008qw}.
For a given GW frequency $h_0^\mathrm{acc}$ is the same as the indirect limit
for ellipticity GW emission, but it is different for a fixed spin period.
Combining with Eq.~(\ref{h0alpha}) we obtain
\begin{eqnarray}
\alpha_\mathrm{acc} &=& \frac{135\sqrt{3}} {2048\pi^{7/2}} F_x^{1/2} P^{7/2} r
M^{-5/4} R^{-9/4} \tilde{J}^{-1}
\nonumber
\\
&=& 5.1\times10^{-6} \left( \frac{F_x} {3.9\times10^{-7}\mbox{ erg\,cm}^{-2}}
\right)^{1/2}
\nonumber
\\
&& \times
\left( \frac{r} {\mbox{2.8 kpc}} \right) \left( \frac{\mbox{800 Hz}} {f}
\right)^{7/2}.
\label{aacc}
\end{eqnarray}
For comparison, \citet{Bondarescu:2007jw} find that $r$-mode evolutions of
accreting neutron stars have $\alpha \approx 10^{-4}$ if there is no run-away
(here the viscosities and thus the parametric instability thresholds differ
from those of young neutron stars).
That nominally corresponds to $P=4$~ms or $f \approx 300$~Hz for Sco~X-1 at the
indirect limit.
Since the $r$-mode amplitudes in these saturation studies are uncertain by at
least a factor of a few due to uncertainties in stellar structure and damping
rates, this is not a precise prediction of the spin period but rather a
statement that Sco~X-1 is consistent with having a short spin period regulated
by $r$-modes at or near equilibrium.
(This was independently found in a more detailed examination of possible
$r$-mode evolutions~\cite{BondarescuWasserman}.)
However non-linear mode evolutions may avoid accretion torque equilibrium
altogether, may spin up as well as down, and may go through intermittent
episodes of $r$-mode activity; and realistic accretion may be more complicated
than what is usually assumed.
Therefore indirect limits from accretion equilibrium are much softer than the
spin-down and age-based limits, which are derived from energy conservation.
(This is a good thing, since the former are much more pessimistic than the
latter~\cite{Owen:2009tj}.)

All-sky surveys for continuous GWs are subject to statistically estimated
indirect limits derived from assumptions about the galactic supernova rate and
distribution of neutron-star parameters at birth.
A simple version due originally to Blandford (unpublished) is described in
\citet{Abbott:2006vg} and is not changed much by $r$-modes since it essentially
relies on $h_0^\mathrm{age}$ and a planar galactic geometry to yield a
population estimate of the brightest signal $h_0^\mathrm{pop}$ a few times
$10^{-24}$ independent of frequency and the poorly-known ellipticity.
Simulations by \citet{Knispel:2008ue} using realistic spatial distributions of
neutron stars produce $h_0^\mathrm{pop}$ values which do depend on ellipticity
and frequency.
Investigating the effect of $r$-mode emission on this type of indirect limit
estimate is a substantial undertaking which I do not attempt here.
However $r$-mode emission has a steeper frequency dependence than ellipticity
emission, resulting in $r$-modes dominating $dE/dt$ for
\begin{equation}
\alpha > \frac{3\sqrt{5}} {16\pi} \frac{I_{zz}} {M\Omega R^3\tilde{J}} \epsilon
= 87\epsilon \left( \frac{P} {\mbox{10 ms}} \right)
\end{equation}
(obtained by equating $I^{2,2}$ to $S^{2,2}$).
This condition is satisfied even by the lowest parametric instability threshold
for the frequencies and ellipticities contributing to the main result (Fig.~5)
of \citet{Knispel:2008ue}.
Thus the presence in a similar simulation of a significant population with
active $r$-modes should spin down stars more quickly into the best frequencies
for GW detection and to produce a higher $h_0^\mathrm{pop}$, even assuming the
scenarios of \cite{Bondarescu:2008qx} and not invoking a low-viscosity
run-away.

The last form of the indirect limit~(\ref{asd}) for known pulsars can also be
used to interpret direct upper limits from all-sky surveys by overlaying the
contours of $\alpha_\mathrm{sd}(f,\dot{f})$ on a plot similar to Fig.~41 of
\citet{Abbott:2007tda}.
Following the discussion around that figure we can say:
The range of the multi-interferometer Hough search from \citet{Abbott:2007tda}
was about 1~kpc at $f=100$~Hz and $|\dot{f}|=10^{-8}$~Hz/s, if $\dot{f}$ for a
star is GW-dominated---and Eq.~(\ref{asd}) tells us that would require
$\alpha\approx0.3$.
A star at $f=1$~kHz with the same $\dot{f}$ would have been detectable up to
100~pc (comparable to the closest known neutron stars) if GW-dominated, and
this would require $\alpha\approx1\times10^{-4}$, which is well within the
range of possibilities for non-linear evolutions.

\section{Discussion}
\label{end}

I have shown that current data analysis methods can detect or set valid upper
limits on continuous GW emission from $r$-modes, in some cases with small
modifications.
I have derived relations needed to interpret GW observational results in terms
of $r$-mode emission.
This allows broad-band searches for continuous GWs to infer $r$-mode amplitudes
or upper limits.

Searches for $r$-modes in known pulsars will also require information on the
ratio of GW frequency to spin frequency \citep{JonesOwen}, and will be more
sensitive to the issue of $r$-mode phase coherence time due to their long
integration times \citep{Wasserman}.

I have also re-derived some commonly-used indirect limits on GW emission for
the case of $r$-modes.
Several young pulsars and non-pulsing neutron stars are interesting targets for
searches for $r$-mode GW emission, and all-sky surveys can have interesting
ranges, even with presently available LIGO and Virgo data.
More young objects further away will be interesting with advanced LIGO and
Virgo.

\acknowledgments

I am grateful for helpful discussions with members of the LIGO and Virgo
Collaborations, especially R. Bondarescu, T. Creighton, and D.~I. Jones; and
with E.~S. Phinney and I. Wasserman.
This work was supported by NSF grant PHY-0855589 and by the LIGO Visitors
Program.
LIGO was constructed by the California Institute of Technology and
Massachusetts Institute of Technology with funding from the National Science
Foundation and operates under cooperative agreement PHY-0107417.
This paper has LIGO Document Number LIGO-P1000019-v3.

\bibliography{rpol}

\end{document}